\documentclass[journal]{IEEEtran}

\usepackage{cite}
\usepackage{amsmath,amssymb,amsfonts}
\usepackage{algorithmic}
\usepackage{graphicx}
\usepackage{textcomp}

\usepackage[colorlinks=true, allcolors=blue]{hyperref}
\usepackage{verbatim}
\usepackage{soul}
\usepackage{mathtools}
\usepackage{booktabs}
\usepackage{siunitx}
\usepackage{multirow}

\usepackage{array}

\soulregister\cite7
\soulregister\ref7
\soulregister\pageref7

\begin{document}

\title{Wave-Equation Migration Velocity Analysis for Multistatic Synthetic Aperture Ultrasound}
\author{Rehman~Ali,                 
        Trevor~M.~Mitcham, 
        Marvin~M.~Doyley,
        Nebojsa~Duric, 
        and~Jeremy~J.~Dahl 
\thanks{Rehman~Ali (Rehman\_Ali@URMC.Rochester.edu) and Nebojsa~Duric (Nebojsa\_Duric@URMC.Rochester.edu) are with Department of Imaging Sciences, University of Rochester Medical Center.} %
\thanks{Trevor~Mitcham (Trevor.Mitcham@ll.mit.edu) is technical staff affiliated with MIT Lincoln Laboratory.} %
\thanks{Marvin~Doyley (m.doyley@rochester.edu) is with the Department of Electrical and Computer Engineering, University of Rochester.} %
\thanks{Jeremy~Dahl (jjdahl@stanford.edu) is with Department of Radiology (Pediatric Radiology) in the Stanford School of Medicine.} %
\thanks{Funding was provided by the National Institute of Biomedical Imaging and Bioengineering (NIBIB) under Grants F32-EB034589 and K99-EB037080.} %
\thanks{Manuscript received April 20, 2026.}} %

\maketitle

\setcounter{figure}{-1}
\begin{figure*}
\includegraphics[width=\textwidth]{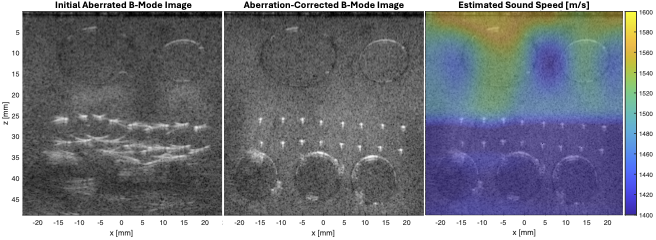}
\caption{\textbf{Graphical Abstract Result. Phantom experiments show dramatic improvements in image quality with measured improvements in point target resolution from 1.22$\pm$1.01 to 0.32$\pm$0.07 mm and lesion contrast from 3.05 to 4.39 dB.}}
\label{fig:graphicalabstract}
\end{figure*}

\begin{abstract}
Sound speed heterogeneities can create aberrations in B-mode ultrasound images by inducing tissue-dependent delays and diffractive effects that conventional beamforming does not incorporate.  By using the Fourier split-step method to simulate pressure fields in heterogenous sound speed media, reverse-time migration (RTM) can reconstruct the B-mode image by cross-correlating transmitted and received pressure fields.  As a result, RTM is differentiable with respect to sound speed.  This enables the reconstruction of the sound speed profile that minimizes the aberration in the B-mode image.  In seismic imaging, this form of diffraction tomography, known as wave-equation migration velocity analysis, can roughly be understood as a type of full-waveform inversion (FWI) that acts in the image domain rather than errors in the received channel data. This is the first work applying WEMVA to medical pulse-echo ultrasound imaging.  Phantom experiments show dramatic improvements in image quality with measured improvements in point target resolution from 1.22$\pm$1.01 to 0.32$\pm$0.07 mm and lesion contrast from 3.05 to 4.39 dB (Graphical Abstract Result—Figure \ref{fig:graphicalabstract}).
\end{abstract}


\section{Introduction}
An accurate, spatially resolved estimate of tissue sound speed is essential for correcting aberration in medical pulse-echo ultrasound imaging. Most existing sound speed estimation methods \cite{podkowa2020convolutional,stahli2020improved,schweizer2023robust,ali2023sound} rely on an approximate linear relationship between the spatial distribution of slowness (the reciprocal of sound speed) and aberration delays or phase differences measured between partial images at each pixel. In these approaches, a path-length matrix encodes the line integrals of slowness along propagation paths and is used to reconstruct the sound speed map via regularized inversion.

Recent advances in differentiable beamforming \cite{simson2025ultrasound,ali2023iterative} make the inherently nonlinear nature of sound speed estimation more explicit. In contrast to conventional methods, which decouple aberration delay estimation from the subsequent inversion, differentiable beamforming defines an objective function based directly on discrepancies between partial images. The gradient of this objective effectively backprojects image-domain errors along the corresponding propagation paths, enabling iterative updates of the sound speed profile via gradient-based optimization.

Despite these advances, most existing approaches for sound speed estimation and distributed aberration correction in B-mode ultrasound rely on ray-based models of time-of-flight within delay-and-sum beamforming frameworks \cite{stahli2020improved,schweizer2023robust,ali2023sound,simson2025ultrasound,ali2023iterative}, typically neglecting refraction and diffraction effects. Although reverse-time migration (RTM) \cite{sava2006time} has been employed for aberration correction \cite{ali2023sound}, it is currently paired with sound speed estimates obtained from ray-based tomography of measured time shifts. This separation leads to an inconsistency in physical modeling: the inversion step neglects wave phenomena, while the imaging step accounts for them.  

In this work, we address this inconsistency by directly differentiating RTM with respect to the sound speed \cite{ali2022open, martiartu2026adjoint, ali2026differentiable, heriard2026physics}, thereby unifying sound speed estimation and aberration correction within a single wave-equation-based framework that fully accounts for refraction and diffraction.  In seismic imaging, this form of diffraction tomography, known as wave-equation migration velocity analysis (WEMVA) \cite{sava2004wave,perrone2014linearized,yang2010wave}, can roughly be understood as a type of full-waveform inversion (FWI) that acts in the image space (i.e., migrated waveforms) rather than the data space (i.e., traces in the received channel data).  This work demonstrate two different forms of WEMVA and its application to medical pulse-echo ultrasound imaging.  The first form of WEMVA minimizes the differences between partial images \cite{ali2022open, perrone2014linearized} produced by single-element transmissions in a multistatic synthetic aperture, or full-matrix capture (FMC), acquisition.  The second form of WEMVA is based on a subsurface-offset extension of RTM \cite{yang2010wave}, whose goal is to drive image content towards zero subsurface offset.

\section{Theory}
\subsection{Nonlinear Least-Squares Formulation}
\label{sec:NLLS}  

Both forms of WEMVA will be formulated as nonlinear least-squares problems:
\begin{equation}
\label{eq:NNLS}
E(\vec{s}) = \frac{1}{2} \left\|\vec{R}(\vec{s})\right\|^2,
\end{equation}
where $\vec{R}$ defines the residual to be minimized as a function of the estimated slowness profile $\vec{s}$.  This nonlinear least-squares formulation enables WEMVA to leverage the same conjugate gradient (CG) algorithm often used in full-waveform inversion (FWI) \cite{ali20242}.  The gradient of the objective function (\ref{eq:NNLS}) is 
\begin{equation}
\label{eq:gradNNLS}
\nabla_{\vec{s}}E := \frac{\partial E}{\partial \vec{s}} = \left(\frac{\partial \vec{R}}{\partial \vec{s}}\right)^H \vec{R}(\vec{s}), 
\end{equation}
where $\left(\frac{\partial \vec{R}}{\partial \vec{s}}\right)^H$ backprojects the residual $\vec{R}(\vec{s})$.  The gradient is computed using reverse-mode automatic differentiation in JAX \cite{bradbury2021jax}.  To implement CG for this objective function (\ref{eq:NNLS}), we also need to calculate the linearized perturbation in the residual $\vec{R}$ for a given perturbation $\Delta\vec{s}_k$ to the slowness: 
\begin{equation}
\label{eq:Perturbation}
\Delta\vec{R} = \left(\frac{\partial\vec{R}}{\partial\vec{s}}\right)\Delta\vec{s}_k
\end{equation}
In JAX, the forward-mode computation (\ref{eq:Perturbation}) of $\Delta\vec{R}$ is implemented using a Jacobian-vector product (JVP). The perturbation $\Delta\vec{R}$ allows us to scale the search direction $\Delta\vec{s}_k$ and update the slowness model $\vec{s}_k$:
\begin{equation}
\label{eq:StepSize}
\alpha_k = -\frac{(\Delta\vec{s}_k)^T\nabla_{\vec{s}}E(\vec{s}_k)}{\|\Delta\vec{R} \|_2^2}
\end{equation}
\begin{equation}
\label{eq:SlownessUpdate}
\vec{s}_{k+1} = \vec{s}_k + \alpha_k \Delta\vec{s}_k
\end{equation}

The formula (\ref{eq:StepSize}) for the step size $\alpha_k$ is analytically exact for linear least-squares problems.  In FWI \cite{ali20242}, this linear approximation is often adequate for convergence.  However, WEMVA can be far more nonlinear than FWI.  In this work, the step size formula (\ref{eq:StepSize}) provides an initial step size for a more complete line search algorithm that consists of bracketing and golden-section search.

\subsection{Preconditioned Conjugate Gradient}
\label{sec:PCG}  

In this work, preconditioning is used to encode prior information, similar to regularization.  Given a symmetric and positive definite preconditioner $\boldsymbol{Q}$, the first iteration of CG applies the preconditioner to the gradient descent direction (i.e., $\Delta\vec{s}_1  = -\boldsymbol{Q}\nabla_{\vec{s}}E(\vec{s}_1)$). Successive iterations update the search direction according to the Polak-Ribière formula: 
\begin{equation}
\label{eq:MomentumCG_PolakRibiere}
\beta_{k} = \frac{\nabla_{\vec{s}}E(\vec{s}_k)^T \boldsymbol{Q}\left(\nabla_{\vec{s}}E(\vec{s}_k)-\nabla_{\vec{s}}E(\vec{s}_{k-1})\right)}{\nabla_{\vec{s}}E(\vec{s}_{k-1})^T \boldsymbol{Q}\nabla_{\vec{s}}E(\vec{s}_{k-1})},
\end{equation}
\begin{equation}
\label{eq:SearchDirectionUpdate}
\Delta\vec{s}_k = -\boldsymbol{Q}\nabla_{\vec{s}}E(\vec{s}_k) + \beta_{k}\Delta\vec{s}_{k-1}.
\end{equation}

Similar to the prior work \cite{ali2023sound}, the symmetric positive definite $\boldsymbol{Q}$ can be modeled as a weighted sum of covariance matrices encoding different priors on the slowness image,
\begin{equation}
\label{covariance}
\boldsymbol{Q} = [(1-\gamma)\boldsymbol{I} + \gamma\boldsymbol{Q_{layer}}]\boldsymbol{Q_{blur}}
\end{equation}
where the blurring operator $\boldsymbol{Q_{blur}}$ is implemented as a convolution with a Hanning window to enforce smoothness, $\boldsymbol{Q_{layer}}$ is implemented as spatial averaging over the lateral dimension to enforce a layered structure in the reconstructed slowness, and $\gamma \in [0,1]$ is the contribution from $\boldsymbol{Q_{layer}}$ relative to the identity operator $\boldsymbol{I}$ after applying $\boldsymbol{Q_{blur}}$.

\subsection{Fourier Split-Step Angular Spectrum Method}
\label{sec:FourierSplitStepMethod}

The frequency response $p_k(x,z,f)$ associated with the $k$th transducer element on the array as a function of location $(x,z)$ and frequency $f$ can be propagated from a depth of $z$ to $z+\Delta z$ using the Fourier split-step method \cite{schwab2018full, jiang2020full, ali2023sound, vyas2012ultrasound}:
\begin{equation}
\label{eq:FourierSplitStep}
\begin{split}
p_k(x,z+\Delta z,f) &= W(x)S(x,z,f)\mathcal{F}^{-1}_{k_x\mapsto x}\\
& H(k_x,z,f)\mathcal{F}_{x\mapsto k_x}p_k(x,z,f),
\end{split}
\end{equation}
\begin{equation}
\label{eq:PropagationFilter}
H(k_x,z,f)=\exp\left(-j2\pi\left((f\bar{s}(z))^2-k_x^2\right)^{1/2} \Delta z\right),
\end{equation}
\begin{equation}
\label{eq:PhaseScreenFunction}
S(x,z,f) = \exp\left(-j2\pi f s_{res}(x,z) \Delta z\right),
\end{equation}
\begin{equation}
\label{eq:LayerwiseAverageSlowness}
\bar{s}(z) = \frac{1}{L}\int_{-L/2}^{L/2} s(x,z) dx,
\end{equation}
\begin{equation}
\label{eq:ResidualSlowness}
s_{res}(x,z)=s(x,z)-\bar{s}(z).
\end{equation}
where $\mathcal{F}_{x\mapsto k_x}$ and $\mathcal{F}^{-1}_{k_x\mapsto x}$ are the forward and inverse Fourier transforms in the lateral dimension ($x$), $W(x)$ is an anti-aliasing window used to suppress wrap-around in the lateral dimension, $s(x,z)$ is the slowness (i.e., reciprocal of sound speed) in the medium as a function of $(x,z)$, and $L$ is the lateral length of the computational domain over which the angular spectrum method is applied.  The boundary condition $\hat{p}_{k}(x,f)=p_{k}(x,z=0,f)$ at $z=0$ represents the frequency response associated with element $k$ at the transducer surface.  

Because this is a Fourier split-step scheme, we decompose $s(x,z)$ into an axially-varying component $\bar{s}(z)$ and a laterally varying component $s_{res}(x,z)$: $H$ is bulk propagation via the angular spectrum method at a slowness $\bar{s}(z)$, and $S$ is a phase screen correction involving the laterally-varying component $s_{res}(x,z)$.  The slowness profile $s(x,z)$ is vectorized as $\vec{s}$. 

\subsection{Target-Following Grid: Adaptive Depth Spacing}
\label{sec:AdaptiveDepth}

On a fixed uniform image grid, image points are placed uniformly in depth according to $z_k = z_0 + k\Delta z$.  However, as $\vec{s}$ changes, an image target initially placed at a depth $z_k$ may appear deeper or shallower.  Ideally, the value of $z_k$ would change with $\vec{s}$ such that it always corresponds to the same image target \cite{ahmed2024spatial,ali2026target}.  This adaptive depth grid would be equally spaced in time rather than depth according to $z_k = z_0 + \sum_{n=1}^{k}{c_n\Delta t}$, where $c_n$ is the reciprocal-average sound speed in layer $n$.  In other words, rather than maintain constant depth-spacing $\Delta z$, we instead calculate a variable $\Delta z = \bar{c}(z)\Delta t$ that maintains constant time-spacing $\Delta t = \bar{s}(z)\Delta z$ (where $\bar{c}(z)=1/\bar{s}(z)$).  The adaptive depth grid significantly mitigates the nonlinearity of the least-squares problem (\ref{eq:NNLS}) and improves the accuracy of the linearized step size formula (\ref{eq:StepSize}).

\subsection{WEMVA Based on Minimizing Image Discrepancies}
\label{sec:WEMVA_ImageDifferences}  

Pulse-echo ultrasound images can be created by propagating transmit and receive wavefields $p_{tx(m)}(x,z,f)$ and $p_{rx(m)}(x,z,f)$ based on the receive channel data collected from pulses emitted by each transmit element $i = 1, ..., N$. Partial images $I_m(x,z)$ can be formed prior to amplitude detection based on the following equation: 
\begin{equation}
\label{eq:ImagingCondition}
\begin{split}
I_m(x,z) &= \int_0^\infty p^{*}_{tx(m)}(x,z,f) p_{rx(m)}(x,z,f) df,
\end{split}
\end{equation}
which represents the frequency-domain implementation of RTM and can be applied to any transmit sequence, including plane-waves \cite{schwab2018full}, focused transmits \cite{ali2021fourier}, diverging waves \cite{ali2023sound}.

This work specifically applies RTM to full-matrix capture (FMC), or multistatic synthetic aperture, data $d_{mn}(t)$, indexed by single-element transmit $m$ and receiver $n$, as a function of time $t$.  In terms of the single-element responses $p_k(x,z,f)$ (\ref{eq:FourierSplitStep}), $p_{tx(m)}(x,z,f)$ and $p_{rx(m)}(x,z,f)$ can be written as:
\begin{equation}
\label{eq:SingleElementTransmit}
\begin{split}
p_{tx(m)}(x,z,f) = p_m(x,z,f), 
\end{split}
\end{equation}
\begin{equation}
\label{eq:ReceivesForGivenTransmit}
\begin{split}
p_{rx(m)}(x,z,f) = \sum_{n=1}^{N} \tilde{d}_{mn}(f) w_n p^*_n(x,z,f).
\end{split}
\end{equation}
where $w_n$ is the apodization over receive elements to maximize the lag-one coherence between images from neighboring transmit elements \cite{ali2023optimal}. Applying (\ref{eq:SingleElementTransmit}) and (\ref{eq:ReceivesForGivenTransmit}) to (\ref{eq:ImagingCondition}) yields: 
\begin{equation}
\label{eq:SummationOverTx}
\begin{split}
I_{m}(x,z) = \sum_{n=1}^{N}\int_0^\infty &p^*_{m}\left(x,z,f\right) \, \tilde{d}_{mn}(f) \\ &w_n \, p^*_{n}\left(x,z,f\right) \, df.
\end{split}
\end{equation}
Moving from (\ref{eq:ImagingCondition}) to (\ref{eq:SummationOverTx}) effectively leverages transmit-receive reciprocity within the multistatic synthetic aperture to reduce the computational and storage costs associated with RTM and its subsurface-offset extension.  One-way responses simulated from each transducer element to each location in space can be used on transmit and receive to synthesize the full two-way response, thereby reducing computational and storage costs to a single set of one-way wavefields rather than separate transmit and receive wavefields for each transmit event.

The first variant of WEMVA involves finding the slowness distribution $\vec{s}$ that minimizes the residual $R_{m}(x,z) = I_{m+1}(x,z)-I_{m}(x,z)$ between the partial images from neighboring single-element transmits $m$ and $m+1$.  In terms of the nonlinear least-squares formulation, $\vec{R}$ is a vectorization of $R_{m}(x,z)$ over $m=1, ..., N-1$, and $(x,z)$ image points.

\subsection{Multistatic Synthetic Aperture B-Mode Imaging}
\label{sec:AberrationCorrectedBModeImaging}  

If we set the receive apodization $w_n=1$, coherent summation over partial images $I_{m}(x,z)$ for each transmit yields the final multistatic synthetic aperture RTM image $I(x,z)$ before amplitude detection and log compression:
\begin{equation}
\label{eq:MultistaticImagingCondition}
\begin{split}
I(x,z) = \sum_{m=1}^{N}\sum_{n=1}^{N}\int_0^\infty &p^*_{m}\left(x,z,f\right) \, \tilde{d}_{mn}(f) \\ &p^*_{n}\left(x,z,f\right) \, df.
\end{split}
\end{equation}
\begin{equation}
\label{eq:decibel}
\begin{split}
I_{display}(x,z) = 20\log_{10}{\left|I(x,z)\right|}. 
\end{split}
\end{equation}

\subsection{WEMVA Based on Subsurface-Offset-Extended RTM}
\label{sec:WEMVA_SubsurfaceOffsetExtension}  

The following equation introduces a subsurface offset $\Delta x$ between the transmitted and received wavefields and is known as the subsurface-offset extension of RTM \cite{yang2010wave}:
\begin{equation}
\label{eq:ExtendedImagingCondition}
\begin{split}
I_m(x,z;\Delta x) = \int_0^\infty &p^{*}_{tx(m)}\left(x+\frac{\Delta x}{2},z,f\right) \\ &p_{rx(m)}\left(x-\frac{\Delta x}{2},z,f\right) \, df.
\end{split}
\end{equation}
Once again, this subsurface-offset extension of RTM is equally applicable to any transmit sequence.  Applying the same subsurface-offset extension to multistatic imaging (\ref{eq:MultistaticImagingCondition}) yields: 
\begin{equation}
\label{eq:MultistaticExtendedImagingCondition}
\begin{split}
I(x,z;\Delta x) = \sum_{m=1}^{N}\sum_{n=1}^{N}\int_0^\infty &p^*_{m}\left(x+\frac{\Delta x}{2},z,f\right) \\ \tilde{d}_{mn}(f) \,\, &p^*_{n}\left(x-\frac{\Delta x}{2},z,f\right) df.
\end{split}
\end{equation}
Note that equations (\ref{eq:ExtendedImagingCondition}) and (\ref{eq:MultistaticExtendedImagingCondition}) simplify to (\ref{eq:ImagingCondition}) and (\ref{eq:MultistaticImagingCondition}), respectively, when subsurface offset $\Delta x=0$.  WEMVA based on the subsurface-offset extension of RTM aims to drive content in the extended image $I(x,z;\Delta x)$ towards $\Delta x=0$ \cite{yang2010wave}.  Therefore, we define the following residual function:
\begin{equation}
\label{eq:SubsurfaceOffsetResidual}
\begin{split}
R(x,z;\Delta x) = W(\left|\Delta x\right|)\,I(x,z;\Delta x), 
\end{split}
\end{equation}
where $W(\cdot)$ is some monotonically increasing function of its input.  We choose $W(x) = \tanh{\left(\frac{x}{a}\right)}$ to gradually taper the gradient as $\Delta x$ increases.  In terms of the nonlinear least-squares formulation, $\vec{R}$ is a vectorization of $R(x,z;\Delta x)$ over subsurface offset $\Delta x$, and $(x,z)$ image points. 

\section{Methods}
\begin{figure*}
\centering
\includegraphics[width=0.98\textwidth]{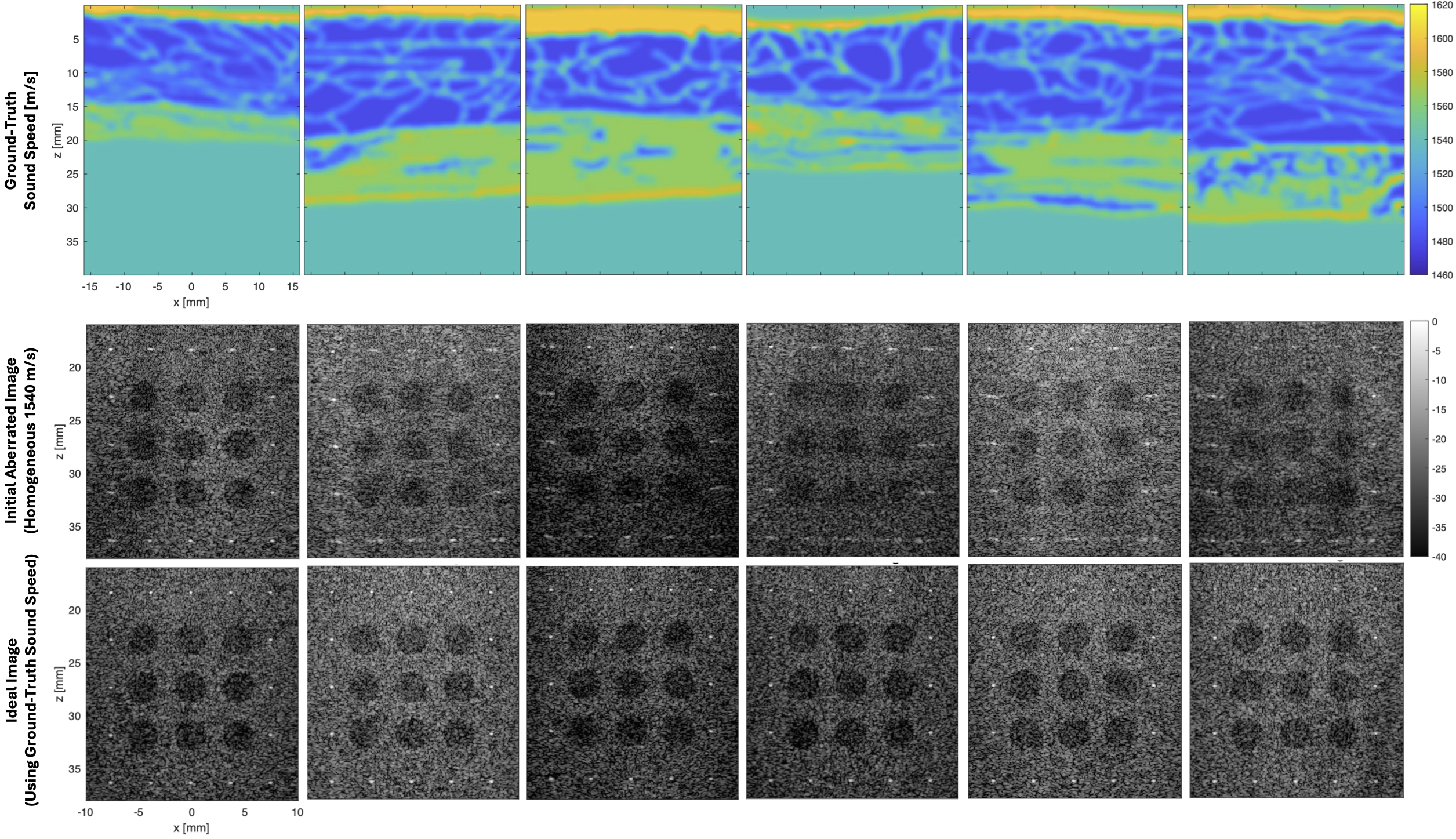}
\caption{Simulations of Aberration Caused by the Sound Speed Heterogeneity of Abdominal Tissue to Test WEMVA.  (Top Row) Six abdominal maps were simulated in k-Wave.  (Middle Row) RTM with an initial homogeneous 1540 m/s sound speed model produces an aberrated B-mode image in each of the six cases.  (Bottom Row) RTM using the true sound speed profile produces the ideal aberration-free B-mode image in each case.}
\label{methods:abdominalmaps}
\end{figure*}

\subsection{k-Wave Simulations in Heterogeneous Media}
\label{sec:simulations}  

Multistatic synthetic aperture channel data simulated in k-Wave \cite{treeby2010k} for the six abdominal models \cite{mast1997simulation} used in prior work \cite{ali2023sound} will be used to compare both forms of WEMVA in terms of resolution and contrast after aberration correction in this work (Figure \ref{methods:abdominalmaps}).  These simulations establish the superiority of the subsurface-offset extension form of WEMVA, so that the remainder of this work focuses on the subsurface-extension formulation of WEMVA.

\subsection{Experimental Data Acquisition}
\label{sec:experiments}  

Hadamard encoded transmit sequences \cite{chiao1997sparse} were used to acquire multistatic synthetic aperture datasets both \textit{in vitro} and \textit{in vivo} on the Verasonics Vantage 256 scanner (Verasonics Inc., Kirkland, WA, USA) using the L12-5 50mm (256 elements; 0.2 mm pitch; 8 MHz center frequency), L12-3v (192 elements; 0.2 mm pitch; 7.5 MHz center frequency), and L7-4 (128 elements; 0.298 mm pitch; 5.5 MHz center frequency) probes.  Phantom experiments were performed in tissue-mimicking gelatin phantoms (using the L12-5 50mm), while \textit{in-vivo} datasets were acquired from the abdomen of obese Zucker rats (using the L12-3v) \cite{telichko2022noninvasive} and a consenting healthy human volunteer (using the L7-4).  Note that the same phantom data used in the prior work \cite{ali2023sound} are reused here.  The data from the healthy volunteer was collected under an institutional review board (IRB) protocol (STUDY00010994) approved by Research Subjects Review Board (RSRB) at the University of Rochester. 

\subsection{WEMVA Parameters}
\label{sec:parameters}  

The main WEMVA parameters of interest are (1) the lateral and axial full-width at half-maximum (FWHM) of the Gaussian blurring kernel $\boldsymbol{Q_{blur}}$, (2) the width $a$ of the subsurface-offset weighting function $W(x) = \tanh{\left(\frac{x}{a}\right)}$, (3) the initial homogeneous sound speed, and (4) the $\gamma$ parameter enforcing the layered structure of the slowness profile.

In all cases except for the phantom experiments, $\boldsymbol{Q_{blur}}$ is a $8\lambda \times 8\lambda$ FWHM blurring kernel, and $a = 4\lambda$, where $\lambda$ is the wavelength at the center frequency of the probe at 1540 m/s.  Due to the sharper discontinuities and stronger aberrations in the phantoms, $\boldsymbol{Q_{blur}}$ and $a$ are enlarged by a factor of 4.5 to $36\lambda \times 36\lambda$ FWHM and $18\lambda$, respectively.  

For each imaging case, the optimal initial homogeneous sound speed $c_{init}\in\{1420, 1460, 1500, 1540, 1580\}$ (in m/s) and the optimal $\gamma \in \{0.9, 0.99, 0.999\}$ was determined by minimizing the final converged value of the objective function (\ref{eq:NNLS}).  In all cases, except for the phantom experiments, $c_{init}$ was determined to be 1540 m/s.  The resulting $\gamma$ values for the k-Wave simulations, \textit{in-vivo} rat experiments, and \textit{in-vivo} data from the healthy human volunteer were 0.999, 0.9, and 0.99, respectively.  In the phantom experiments, $c_{init}$ was determined to be 1460 m/s, and the optimal value of $\gamma$ varied across the three phantoms presented in this work


\subsection{Metrics and Evaluation}
\label{sec:metrics}  
To quantify image quality, we measure the contrast of each hypoechoic lesion relative to its background, and the imaging resolution in terms of the -6 dB lateral beamwidth of each point target. In terms of the mean ($\mu$) and variance ($\sigma^2$) of image values in the target (T) and background (B) regions of interest (ROIs), contrast is defined as
\begin{equation}
\label{definition_contrast}
\text{Contrast} = 20\log_{10}\left(\frac{\mu_T}{\mu_B}\right),
\end{equation}
The background ROI is always chosen to have the same size and imaging depth as the target ROI.

\section{Results}
\begin{table*}
\renewcommand{\arraystretch}{1.2}
\caption{Resolution and Contrast Across B-Mode Images for Each Abdominal Map Shown in Figures \ref{methods:abdominalmaps}, \ref{results:abdominalmaps_imagedifferences}, and \ref{results:abdominalmaps_subsurfaceoffset}}
\label{table:AbdominalMaps}
\centering
\begin{tabular}{c|c|c|c|c|c|c|c}
    Case & Metrics & Map 1 & Map 2 & Map 3 & Map 4 & Map 5 & Map 6 \\
    \hline
    Ideal (Using & -6 dB width [mm]--18 mm depth & 0.20 $\pm$ 0.03 & 0.19 $\pm$ 0.03 & 0.20 $\pm$ 0.03 & 0.20 $\pm$ 0.02 & 0.19 $\pm$ 0.02 & 0.20 $\pm$ 0.03 \\
    Ground-Truth & -6 dB width [mm]--36 mm depth & 0.25 $\pm$ 0.01 & 0.27 $\pm$ 0.02 & 0.26 $\pm$ 0.02 & 0.27 $\pm$ 0.02 & 0.26$ \pm$ 0.01 & 0.27 $\pm$ 0.03 \\
    Sound Speed) & Contrast [dB] & 5.44 & 5.02 & 5.19 & 5.41 & 4.91 & 4.78 \\
    \hline
    Before & -6 dB width [mm]--18 mm depth & 0.32 $\pm$ 0.06 & 0.62 $\pm$ 0.33 & 0.26 $\pm$ 0.06 & 0.41 $\pm$ 0.13 & 0.38 $\pm$ 0.21 & 0.54 $\pm$ 0.38 \\
    Correction & -6 dB width [mm]--36 mm depth & 0.34 $\pm$ 0.16 & 0.25 $\pm$ 0.14 & 0.47 $\pm$ 0.12 & 0.80 $\pm$ 0.57 & 0.77 $\pm$ 0.47 & 0.42 $\pm$ 0.08 \\
    (1540 m/s) & Contrast [dB] & 5.07 & 4.95 & 4.51 & 4.47 & 4.35 & 4.48 \\
    \hline
    After Correction & -6 dB width [mm]--18 mm depth & 0.20 $\pm$ 0.02 & 0.22 $\pm$ 0.03 & 0.21 $\pm$ 0.05 & 0.23 $\pm$ 0.02 & 0.21 $\pm$ 0.02 & 0.21 $\pm$ 0.03 \\
    (Image-Difference & -6 dB width [mm]--36 mm depth & 0.27 $\pm$ 0.02 & 0.35 $\pm$ 0.08 & 0.36 $\pm$ 0.15 & 0.33 $\pm$ 0.06 & 0.36 $\pm$ 0.11 & 0.57 $\pm$ 0.41 \\
    WEMVA) & Contrast [dB] & 5.38 & 4.95 & 5.31 & 5.27 & 5.33 & 5.11 \\
    \hline
    After Correction & -6 dB width [mm]--18 mm depth & 0.20 $\pm$ 0.02 & 0.20 $\pm$ 0.02 & 0.18 $\pm$ 0.02 & 0.19 $\pm$ 0.02 & 0.20 $\pm$ 0.02 & 0.20 $\pm$ 0.03 \\
    (Subsurface-Offset & -6 dB width [mm]--36 mm depth & 0.26 $\pm$ 0.01 & 0.26 $\pm$ 0.01 & 0.26 $\pm$ 0.02 & 0.27 $\pm$ 0.02 & 0.26 $\pm$ 0.01 & 0.26 $\pm$ 0.02 \\
    WEMVA) & Contrast [dB] & 5.45 & 4.98 & 5.18 & 5.26 & 4.92 & 4.67 \\
\end{tabular}
\end{table*}

\begin{figure*}
\centering
\includegraphics[width=0.99\textwidth]{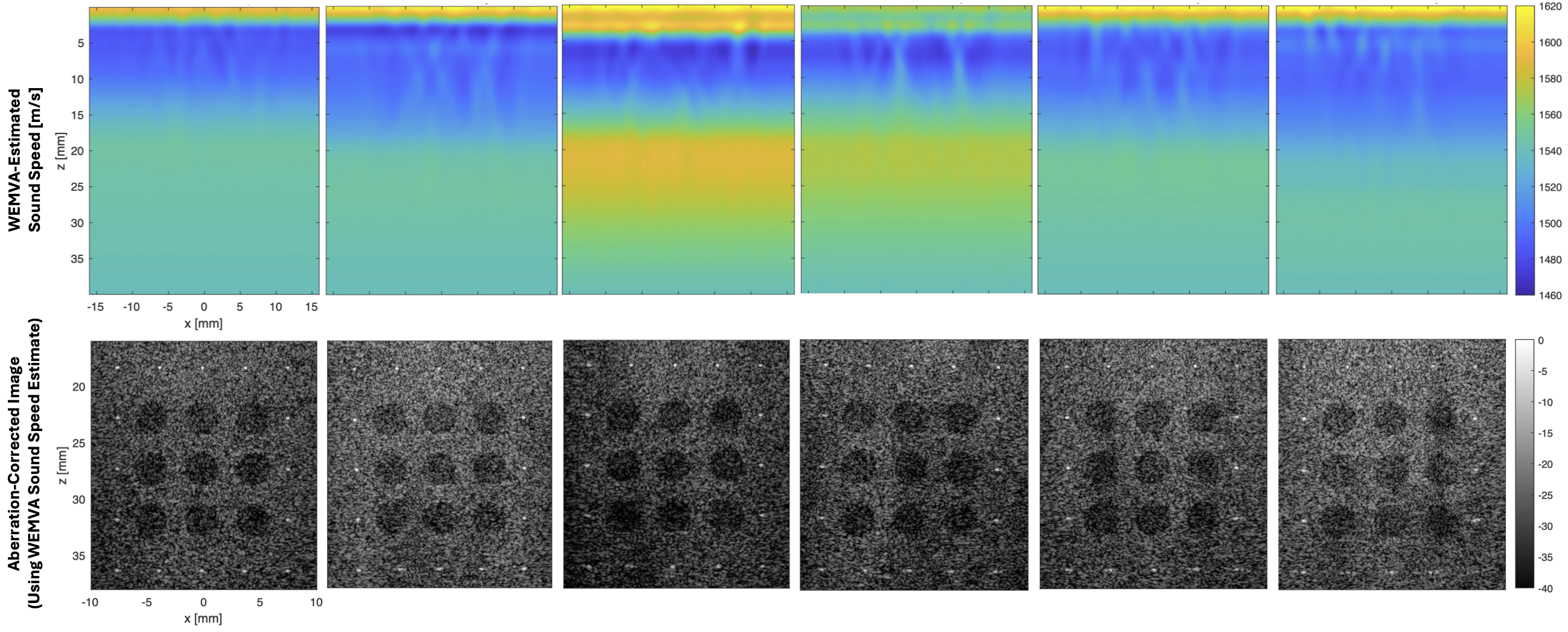}
\caption{WEMVA Based on Minimizing Image Differences in Abdominal Aberration Simulations.  (Top Row) Sound speed estimates for each simulated dataset.  (Bottom Row) Aberration correction using estimated sound speed profile. See Table \ref{table:AbdominalMaps} for the quantification of resolution and contrast.}
\label{results:abdominalmaps_imagedifferences}
\end{figure*}

\begin{figure*}
\centering
\includegraphics[width=0.99\textwidth]{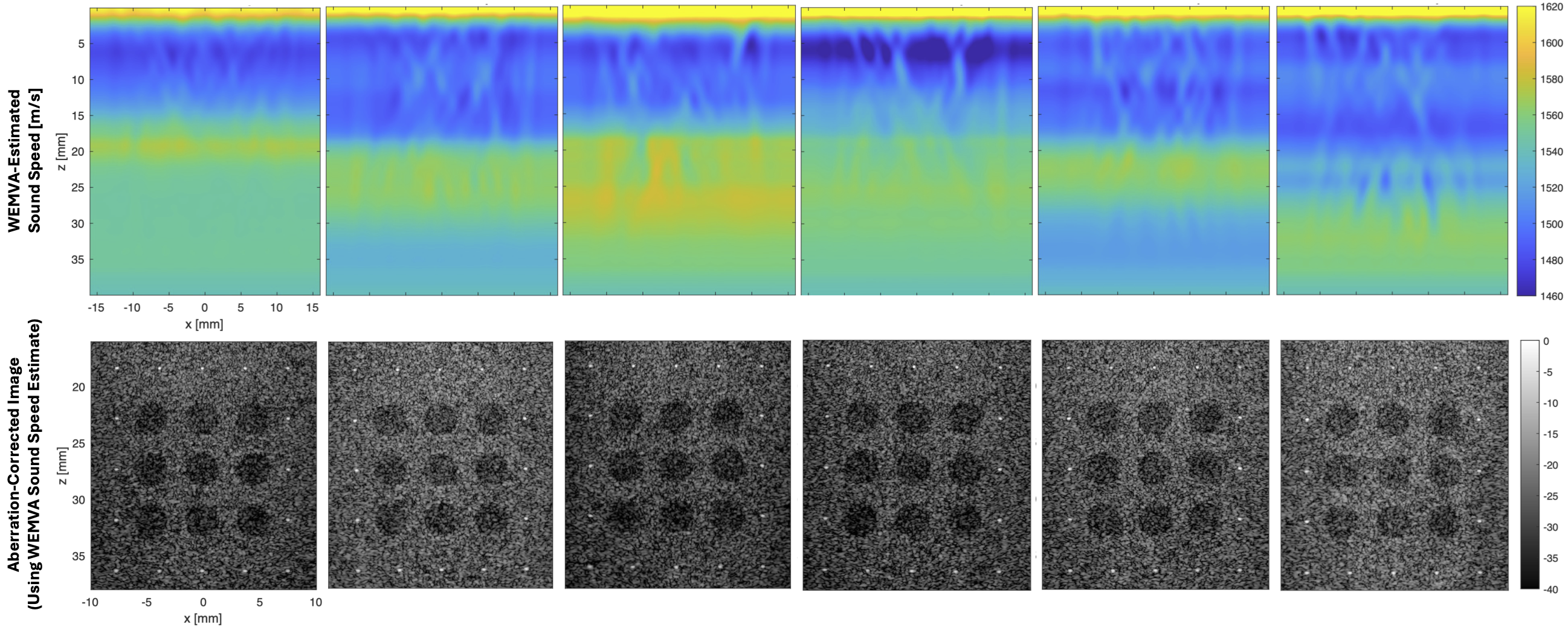}
\caption{WEMVA Based on Subsurface-Offset in Abdominal Aberration Simulations.  (Top Row) Sound speed estimates for each simulated dataset.  (Bottom Row) Aberration correction using estimated sound speed profile. See Table \ref{table:AbdominalMaps} for the quantification of resolution and contrast.}
\label{results:abdominalmaps_subsurfaceoffset}
\end{figure*}

\begin{figure*}
\centering
\includegraphics[width=0.99\textwidth]{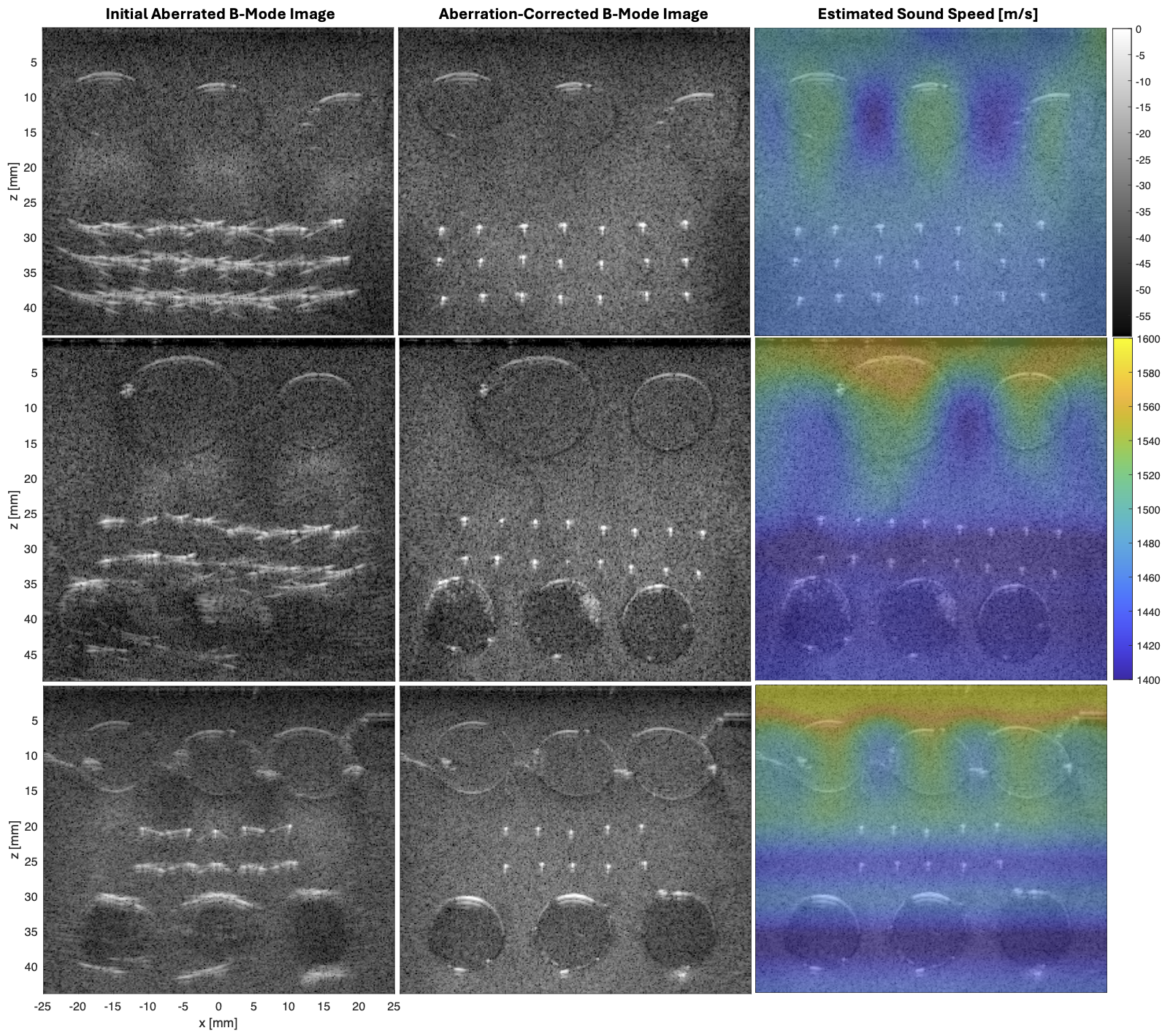}
\caption{Aberration Correction in Phantom Experiments Using Subsurface-Offset WEMVA.  Improvements in point target resolution and lesion contrast are observed in each case.  (Top Row) The optimal value of $\gamma$ was determined to be 0.9 for this phantom. The lateral -6 dB width of the point targets improves from 1.86$\pm$1.31 mm to 0.33$\pm$0.07 mm.  (Middle Row) The optimal value of $\gamma$ was determined to be 0.99 for this phantom. The lateral -6 dB width of the point targets improves from 1.26$\pm$1.10 mm to 0.32$\pm$0.08 mm.  Lesion contrast improves from 0.89 to 2.39 dB.  (Bottom Row) The optimal value of $\gamma$ was determined to be 0.999 for this phantom. The lateral -6 dB width of the point targets improves from 0.61$\pm$0.35 mm to 0.26$\pm$0.07 mm.  Lesion contrast improves from 8.14 to 8.51 dB.}
\label{results:phantoms}
\end{figure*}

\begin{figure*}
\centering
\includegraphics[width=0.99\textwidth]{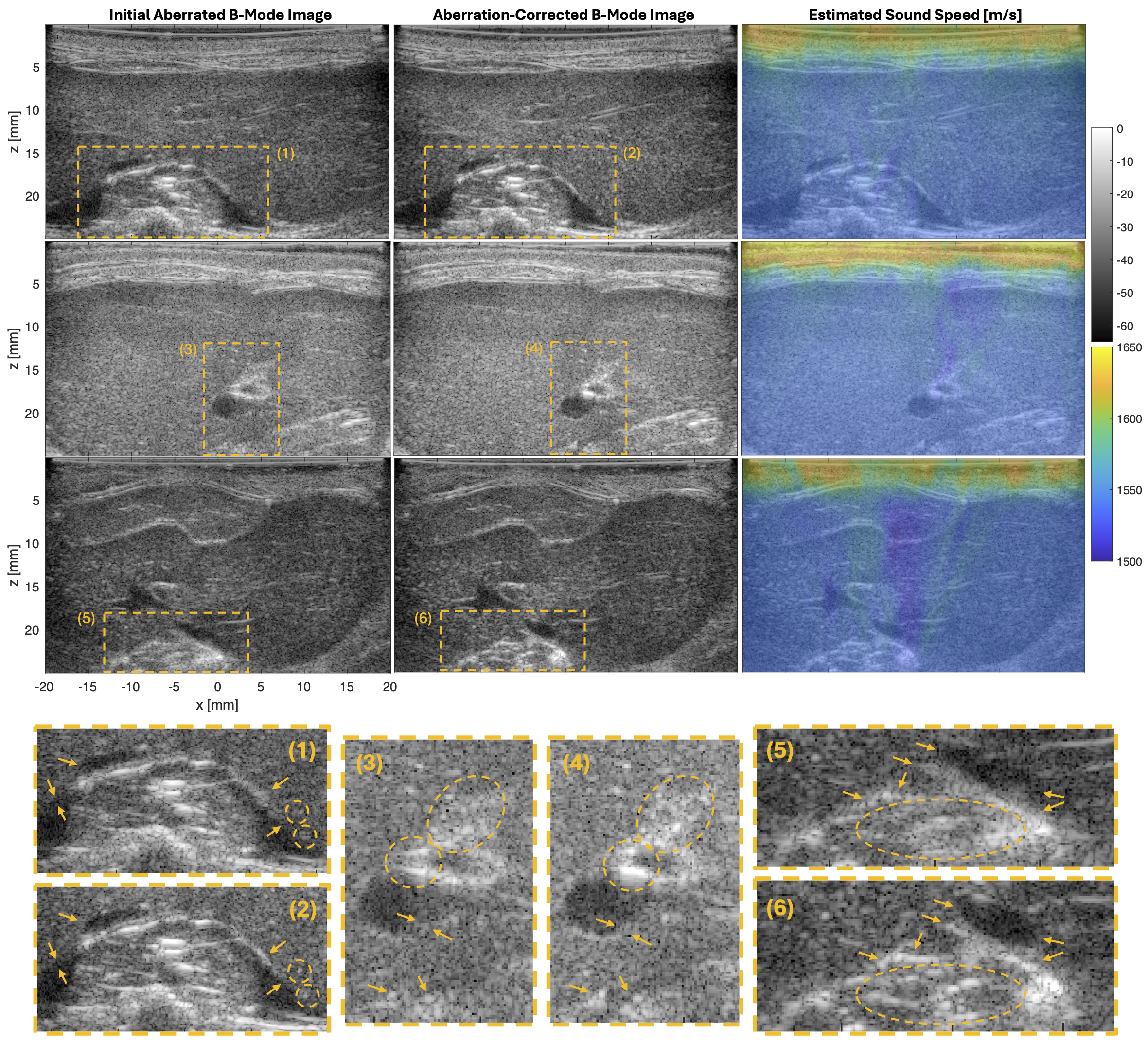}
\caption{Aberration Correction in the Abdomens of Obese Zucker Rats Using Subsurface-Offset WEMVA.  In each case, the focusing of structures at the distal end of the liver improves.  Regions-of-interest (1)-(6) are expanded and annotated to highlight parts of the image where the aberration correction improves image quality. (Top Row) The lateral -6 dB width of the slanted interface to the right of the centermost anechoic space decreases from 0.79 to 0.24 mm.  (Middle Row) The -6 dB edge width measured at the right-edge of the anechoic space decreases from 0.51 to 0.26 mm.  (Bottom Row) The -6 dB edge width measured at the left edge of the anechoic space at the distal end of the liver decreases from 1.09 to 0.36 mm.}
\label{results:ratabdomen}
\end{figure*}

\begin{figure*}
\centering
\includegraphics[width=0.99\textwidth]{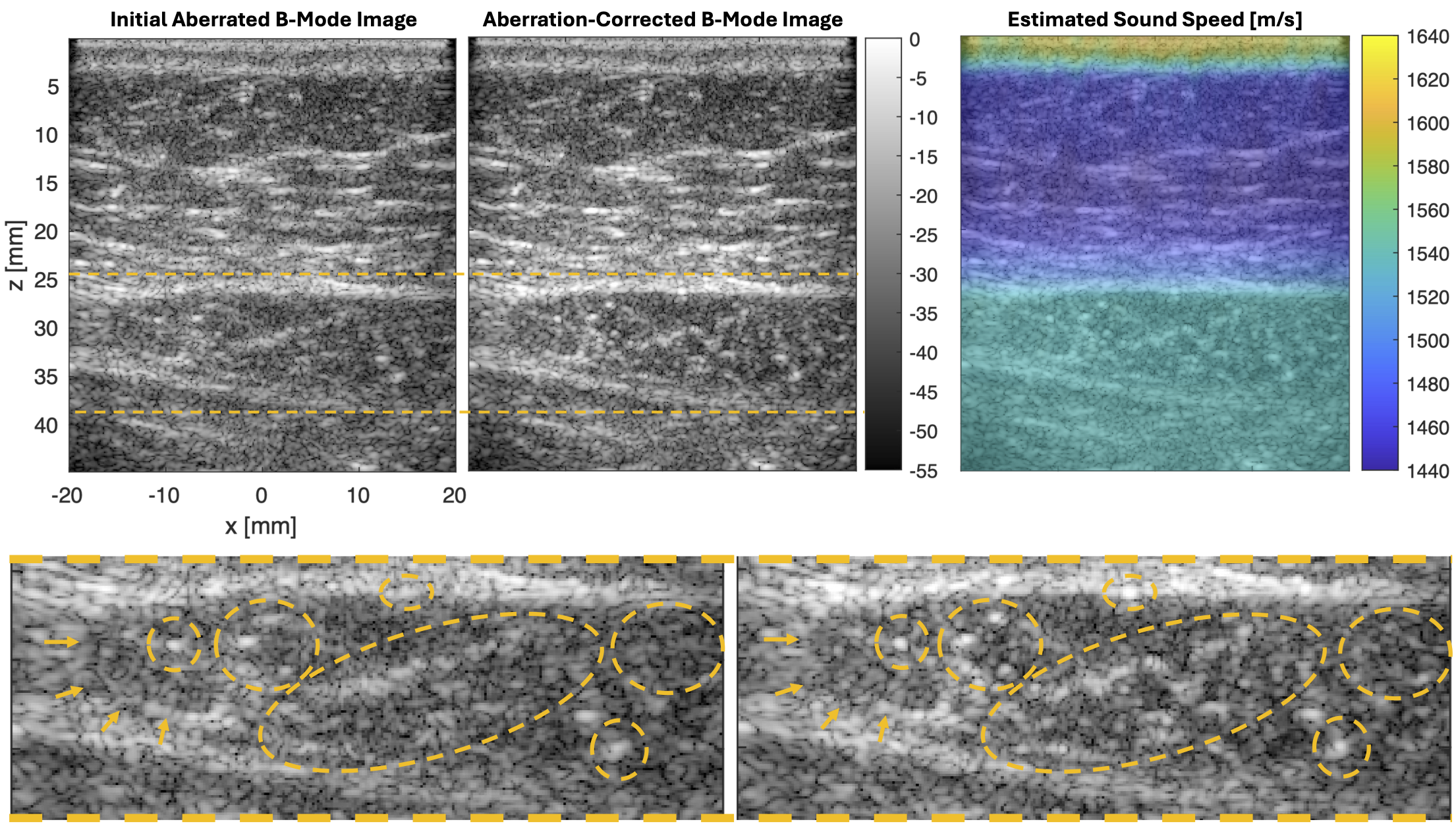}
\caption{Aberration Correction in the Abdominal Wall of a Healthy Human Volunteer Using Subsurface-Offset WEMVA.  The -6 dB widths of point-like structures in the muscle layer of the abdominal wall decreases from 0.88 $\pm$ 0.25 mm to 0.59 $\pm$ 0.17 mm.}
\label{results:superficiallayers}
\end{figure*}

\subsection{In-Silico Validation} 
\label{sec:insilicoresults}

Figures \ref{results:abdominalmaps_imagedifferences}, and \ref{results:abdominalmaps_subsurfaceoffset} show sound speed estimates and aberration-corrected B-mode images for each of the six abdominal aberration simulations shown in Figure \ref{methods:abdominalmaps} using the image-difference and subsurface-offset variants of WEMVA, respectively.  Table \ref{table:AbdominalMaps} reports the -6 dB widths of the point targets at 18 and 36 mm depth and the contrast of the hypoechoic lesions in each B-mode image.  The subsurface-offset variant of WEMVA consistently achieves better spatial resolution (-6 dB target widths) than the image-difference variant.  Additionally, both the spatial resolution and contrast achieved by the subsurface-offset variant of WEMVA are closer to the same measurements in the ideal set of B-mode images using the true sound speed profile.  Additionally, the subsurface-offset WEMVA appears to better capture both lateral and axial variations in sound speed deeper into the medium.  However, neither approach provides strictly quantitative estimates of sound speed because of the velocity-depth ambiguity highlighted in prior work \cite{ali2023sound}.

\subsection{In-Vitro Experiments} 
\label{sec:invitroresults}

Figure \ref{results:phantoms} demonstrates three examples of aberration correction using subsurface-offset WEMVA in phantoms.  There is a progression in the optimal $\gamma$ from 0.9 to 0.99 to 0.999 in the phantom examples shown.  All three cases show significant improvements in point target resolution, and the latter two phantom cases show improvements in lesion contrast and overall image quality. 

In each case, circular inclusions at the top of the phantom have an elevated sound speed relative to the background.  Although the sound speed estimate captures the lateral variations in sound speed introduced by the inclusions, the estimated profile is not strictly quantitative due to non-uniqueness issues caused by the velocity-depth ambiguity; rather, the sound speed estimate represents the components of the true sound speed profile that can be estimated from the measured aberrations, biased by the correlation structure of $\boldsymbol{Q}$.  Increasing $\gamma$ promotes the reconstruction of a more layered sound speed profile via the preconditioner $\boldsymbol{Q}$, which may not be representative of the medium being imaged.  Therefore, the artifactual depth variation in sound speed is a byproduct of the correlation structure imposed on the reconstruction as $\gamma$ increases.  

The final sound speed estimate is also biased by $c_{init}$, as shown in prior work \cite{ali2023sound, ali2023iterative}.  Although $c_{init}$ could be chosen such that it minimizes the objective function, WEMVA does not necessarily converge to a better minimum from this initial guess.  Therefore, we adopted a grid search strategy to find the optimal combination of both $c_{init}$ and $\gamma$ that minimizes the value of the objective function for the final sound speed estimate.  As a result, the optimal $c_{init}$ does not represent the global average sound speed or the sound speed that minimizes the initial value of the objective function; rather, $c_{init}$ represents the initial sound speed that minimizes the final value of the WEMVA objective function.  Unintuitively, the optimal $c_{init}$ may start with a larger initial value of the objective function but may converge to a lower minimum, representing the complexity of the optimization landscape.  

\subsection{In-Vivo Demonstrations} 
\label{sec:invivoresults}

Figure \ref{results:ratabdomen} demonstrates three examples of aberration correction using subsurface-offset WEMVA in the rat abdomen. In each case, the overall image quality and resolution of structures on the distal end of the liver appear to improve. The top row of Figure \ref{results:ratabdomen} corresponds to the dataset previously made open-source as part of the prior work \cite{ali2023sound}.  The middle row of Figure \ref{results:ratabdomen} corresponds to Figure 5 of \cite{simson2025ultrasound}, and the bottom row of Figure \ref{results:ratabdomen} corresponds to Figure 6 of \cite{zhuang2026wavefield}.  In each case, the focusing improvements are comparable to the prior work.  Figure \ref{results:superficiallayers} demonstrates subsurface-offset WEMVA in the superficial layers of the abdomen.  The focusing of structures in the muscle layer appear to improve as a result of the estimated sound speed profile.  The reconstructed sound speed profile appears highly correlated with the layered structure of the tissue.  For example, the skin layer is associated with a high sound speed (1580-1620 m/s) typical of connective tissue, and the region between the muscle and skin has low sound speed (1440-1480 m/s) characteristic of fat.  However, the estimated sound speed inside the muscle layer is unusually low (~1540 m/s) compared to expected ~1580 m/s.  One potential explanation is the lateral variation in the sound speed estimate within the skin layer.  This lateral variation in sound speed produces a lensing effect that compensates for the impact of bulk sound speed errors, similar to near-field biases observed in the prior work as a function of the initial sound speed \cite{ali2023sound}.  

\section{Discussion}
\subsection{Reflection-Mode WEMVA vs Transmission-Mode FWI} 
\label{sec:WEMVAvsFWI}

This work demonstrates the application of WEMVA to medical pulse-echo ultrasound. Both transmission FWI \cite{ali20242} and WEMVA are forms of diffraction tomography \cite{ali2022open} derived from seismic imaging. Transmission FWI reconstructs sound speed by matching simulated and measured transmitted waveforms, with resolution arising from sensitivity to diffraction in forward-scattered data. WEMVA also relies on simulated forward wavefields, but combines transmit and receive wavefields to model the two-way imaging response. Consequently, FWI residuals depend linearly on the simulated wavefield, whereas WEMVA residuals depend bilinearly on transmit-receive products, increasing the nonlinearity of the inverse problem. Furthermore, FWI uses absolute measurements between fixed spatial locations, while WEMVA relies on relative measurements tied to image features with uncertain position \cite{ali2023sound, bickel1990velocity}. As a result, reflection-mode WEMVA is more ill-posed and nonlinear than transmission FWI.  

Although FWI can be applied to reflection data, reflection-mode FWI initially emphasizes impedance contrasts responsible for generating reflections. Starting from a homogeneous model, its first gradient is equivalent to an RTM image \cite{ali2020medical}, and sensitivity to bulk sound speed errors emerges only through later iterations as kinematic inconsistencies between illumination angles accumulate. In contrast, WEMVA directly targets these angle-dependent inconsistencies, making it immediately sensitive to bulk sound speed errors.

\subsection{From Common Midangle to Subsurface Offset} 
\label{sec:CommonMidangle}

Let $I_m(x,z,f;\Delta x)$ denote the integrand of (\ref{eq:ExtendedImagingCondition}):
\begin{equation}
\label{eq:ExtendedImagingConditionBody}
\begin{split}
I_m(x,z,f;\Delta x) = \,\, &p^{*}_{tx(m)}\left(x+\frac{\Delta x}{2},z,f\right) \\ &p_{rx(m)}\left(x-\frac{\Delta x}{2},z,f\right).
\end{split}
\end{equation}
Using plane-wave transmit and receive as in CMA theory \cite{stahli2020improved}:
\begin{equation}
\label{eq:PlaneWaveTX}
p_{tx(m)}\left(x,z,f\right) = \exp\left(-j2\pi\left(k_x^{tx}x+k_z^{tx}z\right)\right),
\end{equation}
\begin{equation}
\label{eq:PlaneWaveRX}
p_{rx(m)}\left(x,z,f\right) = \exp\left(j2\pi\left(k_x^{rx}x+k_z^{rx}z\right)\right),
\end{equation}
\begin{equation}
\label{eq:kTX}
k_x^{tx} = \frac{f}{c}\sin{\theta_{tx}} \qquad k_z^{tx} = \frac{f}{c}\cos{\theta_{tx}}
\end{equation}
\begin{equation}
\label{eq:kRX}
k_x^{rx} = \frac{f}{c}\sin{\theta_{rx}} \qquad k_z^{rx} = \frac{f}{c}\cos{\theta_{rx}}
\end{equation}
where $c$ is the initial beamforming sound speed, and $\theta_{tx}$ and $\theta_{rx}$ are the angles of transmission and reception.  The CMA midangle $\theta_{mid}$ and half-offset angle $\theta_{off}$ are defined as 
\begin{equation}
\label{eq:CMA}
\theta_{mid} = \frac{\theta_{tx}+\theta_{rx}}{2} \qquad \theta_{diff} = \frac{\theta_{tx}-\theta_{rx}}{2}.
\end{equation}

If we insert the definitions (\ref{eq:PlaneWaveTX}) and (\ref{eq:PlaneWaveRX}) of the transmit and receive plane waves into equation (\ref{eq:ExtendedImagingConditionBody}), we obtain:
\begin{equation}
\label{eq:ExtendedImagingConditionPlaneWaves}
I_m(x,z,f;\Delta x) = \exp\left(j2\pi\left(k_x x + k_z z + k_{\Delta x} \Delta x\right)\right),
\end{equation}
\begin{equation}
\label{eq:kx}
k_x = k_x^{tx} + k_x^{rx} = k\sin{\left(\theta_{mid}\right)}, 
\end{equation}
\begin{equation}
\label{eq:kz}
k_z = k_z^{tx} + k_z^{rx} = k\cos{\left(\theta_{mid}\right)}, 
\end{equation}
\begin{equation}
\label{eq:k}
k = \frac{2f}{c}\cos{\left(\theta_{diff}\right)}, 
\end{equation}
\begin{equation}
\label{eq:kDeltax}
k_{\Delta x} = \frac{k_x^{tx} - k_x^{rx}}{2} = \frac{f}{c}\cos{\left(\theta_{mid}\right)}\sin{\left(\theta_{diff}\right)}, 
\end{equation}

The k-space $(k_x,k_z)$ of the zero subsurface-offset ($\Delta x = 0$) image is given by equations (\ref{eq:kx}) and (\ref{eq:kz}), where $(k,\theta_{mid})$ define the polar-coordinate representation of the k-space.  Therefore, a broad range of midangles $\theta_{mid}$ directly translates to angular coverage in k-space and spatial resolution in the reconstructed image.  Note that equation (\ref{eq:k}) for $k$ conveys the same information and is exactly the reciprocal of equation (11) in \cite{stahli2020improved} for the spatial period, consistent with the CMA theory.  Images formed from the same $\theta_{mid}$ but different $\theta_{diff}$ overlap in k-space and are therefore highly correlated. CMA exploits this correlation to estimate sound speed from phase shifts across angular offsets. However, using a broad range of $\theta_{mid}$ improves spatial resolution at the cost of reducing correspondence to well-defined propagation paths, creating a tradeoff between spatial resolution and path specificity.  

This tradeoff is similar to the uncertainty principle from time-frequency analysis in signal processing.  The subsurface-offset extension resolves this tradeoff in a similar manner to the Wigner-Ville distribution in time-frequency analysis \cite{najmi1994wigner}.  For fixed $\theta_{mid}$, different $\theta_{diff}$ values map to different $k_{\Delta x}$ values, so phase alignment across $\theta_{diff}$ becomes equivalent to phase flattening across $k_{\Delta x}$. Since $\Delta x$ and $k_{\Delta x}$ are Fourier duals, this corresponds to concentrating energy at $\Delta x=0$. Thus, subsurface offset achieves the objective of CMA without restricting angular diversity and therefore without sacrificing spatial resolution. It also avoids the CMA assumption that $\theta_{mid}$ and $\theta_{diff}$ are preserved after refraction, since the extended image captures local wavefield behavior directly.

\subsection{Other Transmit Sequences and Imaging Geometries} 
\label{sec:Generalizability}

As discussed in Theory sections \ref{sec:WEMVA_ImageDifferences} and \ref{sec:WEMVA_SubsurfaceOffsetExtension}, RTM and its subsurface-offset extension apply to plane-wave \cite{schwab2018full}, focused \cite{ali2021fourier}, and diverging-wave \cite{ali2023sound} imaging. Existing sound speed estimation methods use different acquisition schemes, including plane-wave \cite{stahli2020improved}, multistatic \cite{simson2025ultrasound}, and virtual-source \cite{schweizer2023robust} synthetic aperture.  While these methods differ in implementation, they all attempt to estimate path-dependent aberration by comparing partially overlapping images.

Plane-wave methods are naturally compatible with CMA, whereas multistatic and virtual-source approaches typically use common midpoint (CMP) transmit and receive subapertures. CMP faces the same tradeoff as CMA: apertures that are too small lose spatial resolution, while apertures that are too large blur path-specific aberration measurements \cite{simson2025ultrasound}. Moreover, CMP depends on far-field approximations and is valid only within limited F-number ranges.  RTM provides a unified framework that models refraction and diffraction directly, and its subsurface-offset extension achieves the objectives of CMA and CMP without sacrificing spatial resolution.

Existing methods also rely on broad unfocused transmissions because ray-based time-of-flight models require wide angular coverage.  Focused transmits are difficult to incorporate because multiple propagation paths intersect near the focus, complicating the attribution of aberration to specific propagation paths. However, this work shows that the subsurface-offset extension of RTM can be applied after summation over transmit and receive, suggesting that focused transmit beams may also be used for sound speed estimation.

The Fourier split-step implementation presented here is limited to linear arrays, but prior work has extended it to curvilinear probes through conformal mapping in polar coordinates \cite{ali2022angular}, and the same subsurface-offset extension applies there.  WEMVA may also be useful in a closed-ring geometry used for transmission FWI \cite{ali20242}, either to regularize FWI or to compare reflection- and transmission-based reconstructions. However, the Fourier split-step method may not easily be adapted to such an imaging geometry and the subsurface offset must extend in both spatial dimensions $(\Delta x, \Delta z)$, increasing computational cost substantially. For this reason, the image-difference formulation of WEMVA with time-domain RTM may be more practical for a closed-ring geometry.

\subsection{Subsurface Offset and the Distortion Matrix} 
\label{sec:DistortionMatrix}

The subsurface-offset extension of RTM is conceptually equivalent to the distortion matrix \cite{lambert2020distortion, cui2026iterative, heriard2026physics}.  Rather than define the subsurface offset $\Delta x$ at a location $x$, the distortion matrix defines input and output coordinates $x_{in}$ and $x_{out}$:
\begin{equation}
\label{eq:ExtendedImagingConditionDistortionMatrix}
\begin{split}
I_m(x_{in},x_{out},z) = \int_0^\infty &p^{*}_{tx(m)}\left(x_{in},z,f\right) \\ &p_{rx(m)}\left(x_{out},z,f\right) \, df.
\end{split}
\end{equation}
\begin{equation}
\label{eq:MultistaticExtendedImagingConditionDistortionMatrix}
\begin{split}
I(x_{in},x_{out},z) = \sum_{m=1}^{N}\sum_{n=1}^{N}\int_0^\infty &p^*_{m}\left(x_{in},z,f\right) \\ \tilde{d}_{mn}(f) \,\, &p^*_{n}\left(x_{out},z,f\right) df.
\end{split}
\end{equation}
with $x_{in}=x+\Delta x/2$ and $x_{out}=x-\Delta x/2$. The optical imaging community describes the same angular-correlation behavior exploited in CMA as the memory effect \cite{lambert2020distortion,osnabrugge2017generalized}. Thus, subsurface offset directly bridges the seismic extended imaging approach to distortion-matrix methods in optics.

\subsection{Complete Extended Reverse-Time Migration} 
\label{sec:ExtendedRTM}

In our prior work \cite{ali2023sound}, we also used the time-lag extension of RTM to estimate aberration delays:
\begin{equation}
\label{eq:ImagingConditionTimeAxis}
\begin{split}
I_m(x,z;\tau) = \int_0^\infty &p^{*}_{tx(m)}(x,z,f) \\ &p_{rx(m)}(x,z,f) \, e^{j2\pi f\tau} \, df \\
= \int_0^\infty &p^{*}_{tx(m)}(x,z,t-\frac{\tau}{2}) \\ &p_{rx(m)}(x,z,t+\frac{\tau}{2}) \, dt, 
\end{split}
\end{equation}
where $\tau$ is the time lag.  This time-lag extended RTM was used to estimate the aberration delay $\Delta\tau = \tau_2-\tau_1$ between partial images $I_{m+1}(x,z;\tau_2)$ and $I_{m}(x,z;\tau_1)$.  The complete extended RTM used in seismic imaging \cite{yang2010wave} combines 2D subsurface offset $(\Delta x, \Delta z)$ and time lag $\tau$:
\begin{equation}
\label{eq:CompleteExtendedImagingConditionDistortionMatrix}
\begin{split}
I_m(x_{in},z_{in},x_{out},z_{out};\tau) = \int_0^\infty p^{*}_{tx(m)}\left(x_{in},z_{in},f\right)& \\ p_{rx(m)}\left(x_{out},z_{out},f\right) \, e^{j2\pi f\tau} \, df& \\
= \int_0^\infty p^{*}_{tx(m)}\left(x_{in},z_{in},t-\frac{\tau}{2}\right)& \\ p_{rx(m)}\left(x_{out},z_{out},t+\frac{\tau}{2}\right) \, dt& \\
\end{split}
\end{equation}
where $(x_{in}, z_{in})=(x+\Delta x/2, z+\Delta z/2)$ and $(x_{out}, z_{out})=(x-\Delta x/2, z-\Delta z/2)$.  This extended formulation closely resembles ADMIRE \cite{byram2015model}, which models contributions between input and output coordinates $(x_{in}, z_{in})$ and $(x_{out}, z_{out})$ to suppress off-axis clutter and reverberation by removing content corresponding to large offsets $(\Delta x, \Delta z)$.  The frequency-domain model used in ADMIRE is achieved using a short-time Fourier transform over time lag $\tau$ in a delay-and-sum beamformer rather than the Fourier split-step method used in RTM.  This suggests that WEMVA and ADMIRE could be combined using the Fourier split-step method within a common extended RTM framework, using WEMVA for aberration correction and ADMIRE for clutter suppression.

\subsection{Role of Regularization} 
\label{sec:Regularization}

The main challenge in sound speed estimation is the velocity-depth ambiguity \cite{bickel1990velocity}, which introduces both ill-conditioning and nonlinearity. Limited angular coverage creates a missing-cone problem \cite{lim2015comparative, gillman2024eliminating}, while the apparent positions of aberrating structures depend on the sound speed profile itself. These effects interact with the bilinearity of WEMVA, producing false minima and poor conditioning.

Progress has come through both improved modeling and regularization. CMA, differentiable beamforming, and RTM improve the fidelity of aberration measurements, while regularization incorporates prior assumptions to compensate for missing angular information. In this work, prior information is imposed through preconditioning, which modifies the search direction directly and accelerates convergence compared with penalty-based regularization. Other approaches include sparsity penalties, TV regularization, region-based priors, and learned regularization \cite{rau2021speed,beuret2026total,stahli2020bayesian,yolgunlu2024learned}. Because regularization strongly influences the reconstruction, the resulting sound speed estimates are not fully quantitative.

Optimization strategy also affects regularization performance. Although this work uses least-squares formulations with closed-form step size initialization, the nonlinearity of WEMVA may favor quasi-Newton methods such as the Limited-memory Broyden–Fletcher–Goldfarb–Shanno (L-BFGS) \cite{nocedal1999numerical} algorithm.  More sophisticated objectives, such as the ratio of norms described in adaptive waveform inversion (AWI) \cite{warner2016adaptive,guasch2019adaptive,yong2023localized}, may better utilize subsurface offset.  

\subsection{Computational Cost} 
\label{sec:ComputationalCost}

The Fourier split-step method enables efficient RTM through FFT-based propagation. On an NVIDIA RTX PRO 6000 Blackwell graphics processing unit (GPU), the full RTM reconstruction in Figure \ref{results:phantoms} requires 2 seconds, while the corresponding subsurface-offset WEMVA gradient requires 12 seconds. The layered propagation structure naturally accumulates aberration with depth and provides an efficient mechanism for backprojecting image aberrations onto the slowness model.

The main computational burden is memory usage. WEMVA requires forward wavefields to be available during the reverse pass for gradient computation. Storing all wavefields is prohibitively expensive, while recomputing them incurs overhead. Checkpointing balances this tradeoff by storing wavefields at selected depths and recomputing intermediate states as needed \cite{anderson2012time}. This checkpointing strategy is implemented in JAX \cite{bradbury2021jax} and substantially reduces memory demands.

Using subsurface offset also reduces residual dimensionality compared with CMA or CMP. Those methods require many partial images before phase comparisons can be formed, generating residuals with a large number of entries. Subsurface-offset imaging eliminates this intermediate step and directly forms a compact extended image, reducing both memory cost and computational complexity.

\section{Conclusions}
This work demonstrates differentiable RTM, or WEMVA, for sound speed estimation and aberration correction in handheld pulse-echo ultrasound. As a form of image-domain FWI, WEMVA iteratively updates the sound speed profile to minimize aberration in the reconstructed image. Unlike prior differentiable beamforming approaches, the RTM framework models both refraction and diffraction, enabling more accurate aberration modeling in heterogeneous media. We show that the subsurface-offset formulation of WEMVA provides a more effective residual than image-difference formulations by directly measuring image consistency in an extended image domain while avoiding the need to construct large sets of partial images. This formulation generalizes prior CMA and CMP approaches within a unified full-wave framework, preserving spatial resolution while improving sensitivity to path-dependent aberration. \textit{In-silico}, \textit{in-vitro}, and \textit{in-vivo}  experiments demonstrate substantial gains in overall B-mode image quality. These results establish subsurface-offset WEMVA as a practical and theoretically advantageous framework for reflection-mode sound speed estimation and aberration correction in pulse-echo ultrasound. All code and datasets used in this work have been made publicly available at 
\url{https://github.com/rehmanali1994/WEMVA} (DOI: \url{https://doi.org/10.5281/zenodo.19906865}).

\section*{Acknowledgment}
The authors would like to thank Dr. Naiara Korta Martiartu and her PhD student Chenyu Cui at the Ecole polytechnique fédérale de Lausanne (EPFL) for conversations that helped draw the connection between the subsurface-offset extension described in this work and the distortion matrix method used in their work \cite{cui2026iterative}.  The authors also thank Dr. Mohammad Mehrmohammadi and clinical research coordinators Laurie Christensen and Michael Doud, all from the Department of Imaging Sciences at the University of Rochester Medical Center, for their assistance with the IRB protocol and the informed consent process.

\bibliography{references}{}
\bibliographystyle{IEEEtran}

\end{document}